\documentclass[twocolumn,showpacs,preprintnumbers,amsmath,amssymb]{revtex4}

\usepackage{color,latexsym,epsfig,amssymb,amsfonts,amsmath,graphicx,bbm,enumerate,bm}
\usepackage{graphicx}
\usepackage{epstopdf}
\usepackage{amsfonts}
\usepackage{amssymb}

\newcommand{\ket}[1]{\left | \, #1 \right \rangle}
\newcommand{\bra}[1]{\left \langle #1 \, \right |}
\newcommand{\braket}[2]{\left\langle\, #1\,|\,#2\,\right\rangle}
\newcommand{\mi}{\mathrm{i}}
\newcommand{\rmd}{\mathrm{d}}

\begin{document}

\title{Theory of noise suppression in $\Lambda$-type quantum memories by means of a cavity}

\author{J.~Nunn$^1$, S.~Thomas$^{1,2}$, J.~H.~D.~Munns$^{1,2}$, K.~T.~Kaczmarek$^1$, C.~Qiu$^{1,3}$, A.~Feizpour$^{1}$, E.~Poem$^{1}$, B.~Brecht$^{1}$, D.~J.~Saunders$^{1}$, P.~M.~Ledingham$^1$, Dileep~V.~Reddy$^4$, M.~G.~Raymer$^4$, I.~A.~Walmsley$^1$}
\affiliation{$^{1}$Clarendon Laboratory, University of Oxford, Parks Road, Oxford OX1 3PU, United Kingdom\\
$^{2}$ QOLS, Blackett Laboratory, Imperial College London, London SW7 2BW, UK\\
$^{3}$Department of Physics, Quantum Institute for Light and Atoms, State Key Laboratory of Precision Spectroscopy, East China Normal University, Shanghai 200062, People's Republic of China\\
$^{4}$Oregon Center for Optics, Department of Physics, University of Oregon, Eugene, Oregon 97403, USA
}

\begin{abstract}
Quantum memories, capable of storing single photons or other quantum states of light, to be retrieved on-demand, offer a route to large-scale quantum information processing with light. A promising class of memories is based on far-off-resonant Raman absorption in ensembles of $\Lambda$-type atoms. However at room temperature these systems exhibit unwanted four-wave mixing, which is prohibitive for applications at the single-photon level. Here we show how this noise can be suppressed by placing the storage medium inside a moderate-finesse optical cavity, thereby removing the main roadblock hindering this approach to quantum memory.
\end{abstract}


\maketitle
\section{Introduction}
The need for low-loss active switching or synchronisation of non-deterministic operations in a linear-optical quantum information processor has emerged over the past few years as a \emph{sine qua non} for the development of large-scale photonics-based quantum technologies \cite{Migdall:2002cr,McCusker:2008ve,Nunn:2013aa,Francis-Jones:2015aa,Rohde:2015aa}. Experiments with optical switches are making rapid progress \cite{Makino:2015aa,Kaneda:2015aa,Meany:2014aa,Mendoza:2015aa}. Quantum memories based on the coherent and reversible absorption of photons in an atomic ensemble are being developed by many groups as an alternative to optical switching \cite{Bussieres:2013aa,Afzelius:2010aa,Hammerer:2010vn,Simon:2010aa,Lvovsky:2009ve}, with impressive demonstrations of the preservation of quantum correlations and of temporal synchronisation of photons, using cold atoms \cite{Chen:2006aa} and cold doped crystals \cite{Bussieres:2014aa}. An important class of memory protocol is based on stimulated two-photon transitions in a $\Lambda$-type atomic ensemble, where a bright control laser field couples the incident signal photons to a ground-state coherence in the atoms \cite{Reim:2010kx}. Memories based on electromagnetically-induced transparency (EIT) \cite{Fleischhauer:2002ph,Choi:2008xi,Kupchak:2015aa} and on far-off-resonant Raman absorption \cite{Nunn:2007aa,Ding:2015aa,Hosseini:2011zr} both fall into this category, and in the following we will refer to all such memories as \emph{$\Lambda$-memories}. $\Lambda$-memories in cold atoms have successfully stored single photons, but at room-temperature it was found that fluorescence noise \cite{Manz:2007aa} and four-wave mixing \cite{Phillips:2008uq} became problematic. Our group recently interfaced a single-photon source with a Raman memory and measured the photon number statistics of the retrieved fields \cite{Michelberger:2015aa}. There it was found that while fluorescence noise was negligible for off-resonant storage of short pulses, the thermal noise contributed by four-wave mixing destroyed the anti-bunching characteristic of single photons. Four-wave mixing noise is therefore the key roadblock preventing the implementation of $\Lambda$-memories at room temperature \cite{Lauk:2013aa,Dabrowski:2014aa}. A number of solutions have been suggested \cite{Romanov:2015aa,Zhang:2014aa}. We recently demonstrated that four-wave mixing can be suppressed by placing the memory storage medium inside an optical cavity \cite{Saunders:2015ab}. In this paper we present a theoretical analysis of this method of noise suppression. We show that a moderate-finesse cavity can, under realistic conditions, suppress four-wave mixing noise to a negligible level, while at the same time dramatically reducing the footprint and energy requirements of the memory compared to free-space implementations, and maintaining broadband operation. We also point out that the cavity-enhanced memory is nearly perfectly-temporal-mode-selective \cite{Reddy:2015aa}, making it an appealing system for chronocyclic encodings of quantum information \cite{Brecht:2015ac}.

\section{Four wave mixing in $\Lambda$-memories}
\begin{figure}[h]
\begin{center}
\includegraphics[width=\columnwidth]{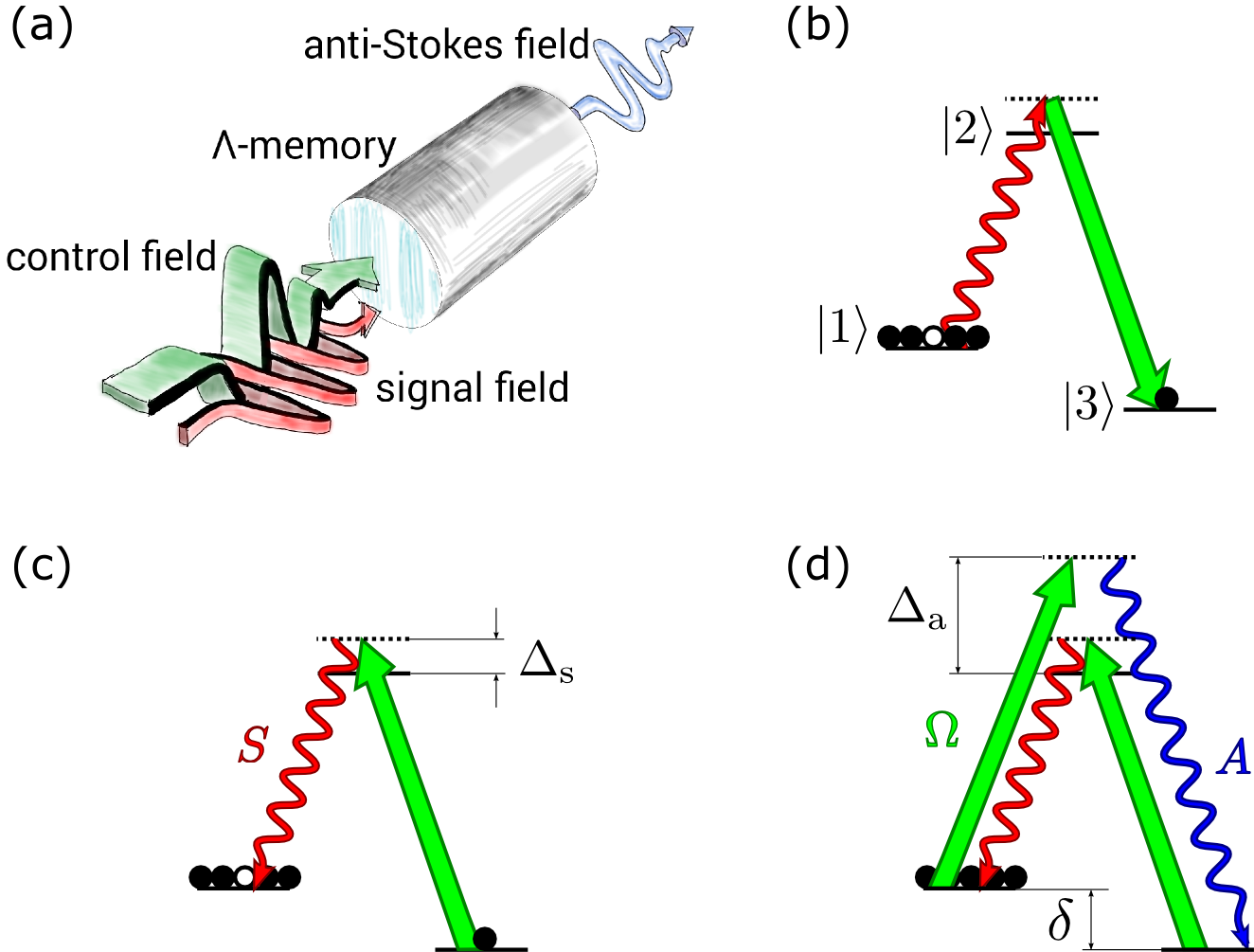}
\caption{(a) A control field (green) mediates the storage of a signal field (red) in a free-space $\Lambda$-memory. The control field can drive the spontaneous Raman emission of an anti-Stokes field (blue) that generates a spurious excitation in the memory. (b) Atomic transitions driven in the storage interaction (c) the retrieval interaction. (d) Four-wave mixing noise refers to the retrieval of the spurious excitation generated by the anti-Stokes scattering.}
\label{fig:mem_and_FWM}
\end{center}
\end{figure}

Although the detailed dynamics in $\Lambda$-memories differ depending on the protocol employed (EIT, Raman, $\Lambda$-GEM), all such memories share the same kinematical description, shown in Fig.~\ref{fig:mem_and_FWM}. That is, an ensemble of $\Lambda$-type atoms are prepared in one of their two ground states, and then an incident signal field is coupled to the empty storage state by a strong control field tuned into two-photon resonance with the signal. Four-wave mixing arises in this system when the strong control field couples to the ground state and drives off-resonant spontaneous Raman scattering (in the diagram this is anti-Stokes scattering because we have chosen to pump the atoms into the higher of the two ground state levels). The scattered anti-Stokes field is at a different frequency from the signal field and does not directly contribute any noise. But each scattering event is accompanied by one of the atoms switching state from $\ket{1}$ to $\ket{3}$. That is, the scattering spontaneously generates excitations of the ground-state coherence that are indistinguishable from the excitations produced by successful storage of the signal field. When the memory is read-out, these excitations are retrieved as noise. The nomenclature `four-wave mixing' applies because there are four optical fields that coherently interact, even though they may not overlap in time. These are: the control; the anti-Stokes; the control again and finally the retrieved (noisy) signal.

\section{Model}
\begin{figure}[h]
\begin{center}
\includegraphics[width=5.5cm]{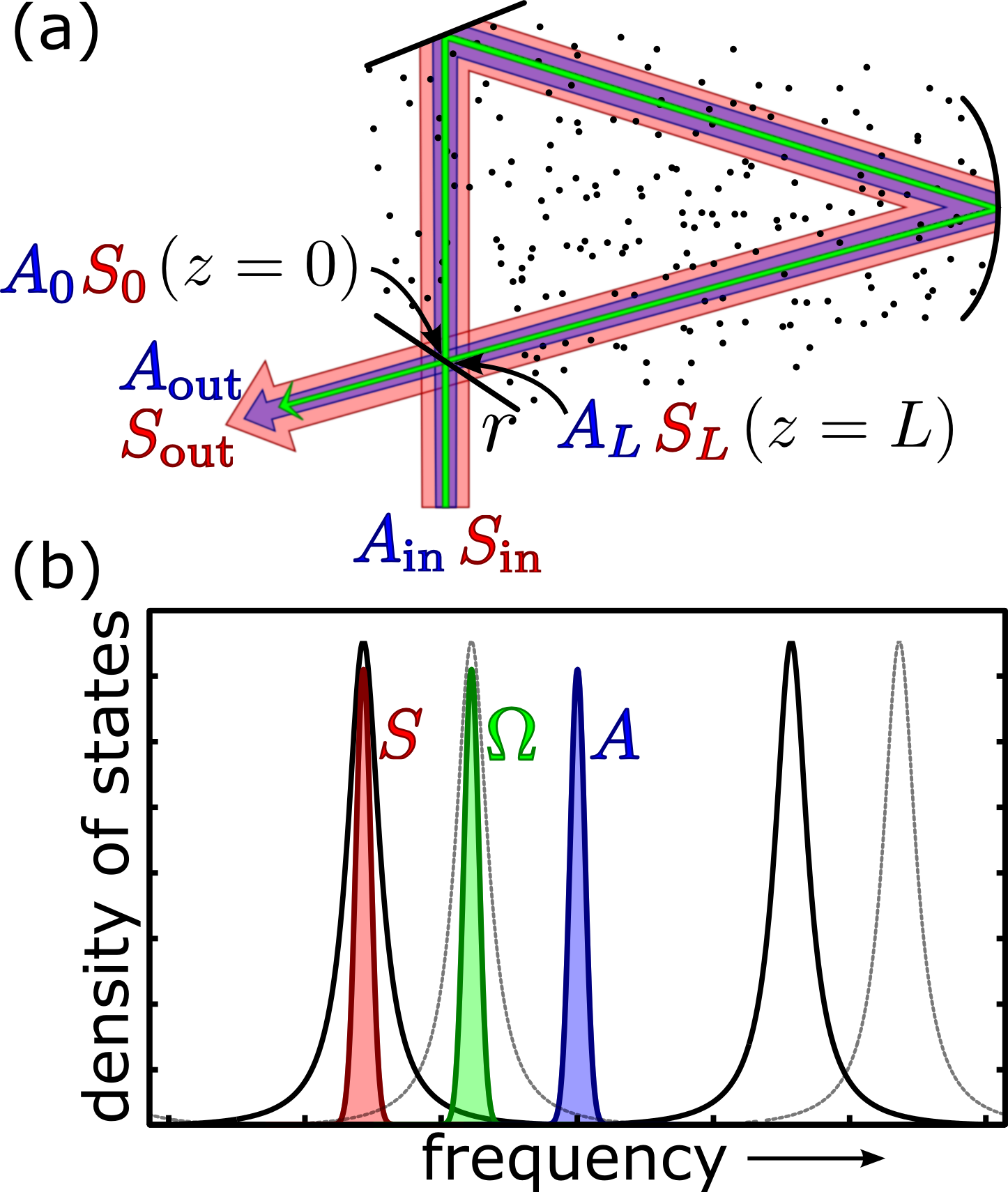}
\caption{(a) We model the dynamics of the cavity-enhanced $\Lambda$-memory by considering the propagation of Stokes and anti-Stokes fields around a ring cavity through a storage medium comprising an ensemble of $\Lambda$-type atoms. The boundary conditions imposed by the input/output coupler with amplitude reflectivity $r$ connects the intra-cavity fields to the incident and emitted fields. (b) The cavity resonances are tuned to suppress the density of scattering states (black peaks) at the anti-Stokes frequency with respect to the Stokes frequency. Tuned mid-way between the Stokes and anti-Stokes frequencies, the control field can be resonantly coupled into the cavity in an orthogonal polarisation mode (grey peaks) \cite{Saunders:2015aa}.}
\label{fig:cavity_memory}
\end{center}
\end{figure}

We now propose and analyse a scheme to inhibit four-wave mixing by using an optical cavity to modify the density of scattering states so that Stokes scattering is enhanced and anti-Stokes scattering is suppressed. That is, we tune the cavity into resonance with the Stokes frequency --- this is also the frequency of the signal to be stored --- whereas we ensure that the anti-Stokes frequency is anti-resonant (Fig.~\ref{fig:cavity_memory} (b)). Below we introduce a model that shows how four-wave mixing noise is suppressed by this arrangement.

We consider a $\Lambda$-memory storage medium with ground state splitting $\delta$ placed inside a ring cavity \footnote{The ring geometry simplifies the analysis to follow, but it also has the practical advantage that the intra-cavity fields are running waves without nodes at zero intensity. This ensures that atoms interacting with the fields during read-in do not escape interaction at retrieval by diffusing into the `dark' field nodes during the storage time.} as shown in Fig.~\ref{fig:cavity_memory} (a). Such a cavity can be successfully described with cavity input-output theory \cite{Walls:2007aa} and the operation of $\Lambda$-memories in a cavity has been analysed in this way \cite{Gorshkov:2007aa}. However, cavity input-output theory is not suited to the description of fields tuned out of resonance with a cavity. To proceed, we instead follow the generalised input-output theory of Raymer and McKinstrie \cite{Raymer:2013aa} and consider the traveling-wave propagation of the signal (Stokes) field $S$ and the anti-Stokes field $A$ around the ring-cavity. The fields interact with the atomic ensemble in the presence of the control pulse, with Rabi frequency $\Omega$, according to the linearised Maxwell-Bloch equations, which in the limit of a sufficiently smooth control, such that the excited state $\ket{2}$ can be adiabatically eliminated, take the form \cite{Phillips:2011aa,Wu:2010aa,Wasilewski:2006th}
\begin{eqnarray}
\nonumber \left(c\partial_z + \partial_t\right)S&=&\mi c\sqrt{\frac{d\gamma}{L}}\frac{\Omega}{\Gamma_\mathrm{s}} B -\kappa_\mathrm{s}S,  \\
\nonumber \left(c\partial_z + \partial_t\right)A&=&\mi c\sqrt{\frac{d\gamma}{L}}\frac{\Omega}{\Gamma_\mathrm{a}} B^\dagger -\kappa_\mathrm{a}A,  \\
\nonumber \partial_t B &=& -\mi \sqrt{\frac{d\gamma}{L}}\frac{\Omega^*}{\Gamma_\mathrm{s}}S + \mi \sqrt{\frac{d\gamma}{L}}\frac{\Omega}{\Gamma_\mathrm{a}}A^\dagger \\
\label{Maxwell}& & - \left[ \frac{1}{\Gamma_\mathrm{s}}+\frac{1}{\Gamma_\mathrm{a}^*}\right]|\Omega |^2  B,
\end{eqnarray}
where the $z$-coordinate parameterises the position along the folded optical path inside the cavity. The system (\ref{Maxwell}) is to be interpreted as describing the evolution of the slowly varying annihilation operators $S$, $A$ in the Heisenberg picture, with the bosonic spin wave annihilation operator given by $B = \sum_{j \in [z,z+\delta z]}\ket{1}_{\!j}\!\!\bra{3}/\delta z \sqrt{N/L}$ describing the amplitude of the Raman coherence excited in the $\Lambda$-memory. The optical and spin wave fields satisfy canonical commutation relations $[S(t,z),S^\dagger(t',z)]=[A(t,z),A^\dagger(t',z)]=\delta(t-t')$, $[B(t,z),B^\dagger(t,z')]=\delta(z-z')$. The operator nature of (\ref{Maxwell}) is key to the analysis of spontaneous four-wave mixing noise. The meanings of the other symbols are as follows. $\Gamma_\mathrm{s,a} = \gamma - \mi \Delta_\mathrm{s,a}$ denotes the complex detuning of the signal and anti-Stokes fields from the atomic resonance with homogeneous linewidth $\gamma$. The complex decay rates $\kappa_\mathrm{s,a}$ account for dispersion, absorption and other scattering losses as the fields propagate. Strictly the losses in (\ref{Maxwell}) should be accompanied by Langevin noise operators which maintain the bosonic commutation relations of the field operators, but vacuum noise does not contribute to the signal intensity and can be neglected \cite{Gorshkov:2007aa}. The signal coupling strength is parameterised by the single-pass resonant optical depth $d$ of the atomic ensemble, with $N$ atoms in the signal beam path of length $L$ \cite{Gorshkov:2007qm}. Neglecting inhomogeneous broadening (generally valid far from resonance) the absorption and dispersion are given by 
\begin{equation}
\label{kappas}
\kappa_\mathrm{s} = \frac{d\gamma}{\tau\Gamma_\mathrm{s}}\,;\qquad \kappa_\mathrm{a}=\frac{d\gamma}{\tau\Gamma_\mathrm{a}^+},
\end{equation}
where $\Gamma_\mathrm{a}^+ = \gamma-\mi (\Delta_\mathrm{a}+\delta)$ is the complex detuning of the anti-Stokes field from the populated transition and $\tau = L/c$ is the cavity roundtrip time. However in an experiment the roundtrip absorption and loss can be inferred from the cavity transmission spectrum \cite{Munns:2015}. Note that we have neglected decoherence of the spin wave, since this is by assumption slow on the time-scale of the memory interactions. The system (\ref{Maxwell}) can be solved analytically \cite{Wu:2010aa}, but in the limit that the interaction with the atoms is weak over the course of a single pass through the cavity, the fields $A_L$, $S_L$ emerging from the interaction at $z=L$ can be related to the amplitudes $S_0$, $A_0$, $B_0$ at $z=0$ by a Taylor expansion,
\begin{eqnarray}
\nonumber S_L &\approx& S_0 + L \partial_z S |_{z=0} \\
\nonumber &=& e^{-\kappa_\mathrm{s}\tau}S_0 + \mi c\tau \sqrt{\frac{d\gamma}{L}} \frac{\Omega}{\Gamma_\mathrm{s}} B_0 -\tau \partial_t S_0;\\
\nonumber A_L &\approx& A_0 + L \partial_z A |_{z=0}\\
\label{Taylor} &=& e^{-\kappa_\mathrm{a}\tau}A_0 + \mi c\tau \sqrt{\frac{d\gamma}{L}} \frac{\Omega}{\Gamma_\mathrm{a}} B_0^\dagger -\tau \partial_t A_0,
\end{eqnarray}
where we assumed $|\kappa_\mathrm{s,a}|\tau \ll 1$ in forming the exponentials. 

To capture the intra-cavity dynamics we now `close the loop' by imposing as a boundary condition the beam splitter relation at the input-output coupler, with amplitude reflectivity $r$ (assumed real for simplicity),
\begin{eqnarray}
\nonumber S_0 &=& re^{\mi k_\mathrm{s} L}S_L + t_rS_\mathrm{in},\\
\label{bound} A_0 &=& re^{\mi k_\mathrm{a} L}A_L + t_rA_\mathrm{in},
\end{eqnarray}
where $t_r = \sqrt{1-r^2}$ is the input-output coupler amplitude transmission, $k_\mathrm{s}$ ($k_\mathrm{a}$) denotes the signal (anti-Stokes) carrier wavevector, and we have introduced the input field amplitudes $A_\mathrm{in}$, $S_\mathrm{in}$ impinging on the cavity from the outside. Substituting the solutions (\ref{Taylor}) into (\ref{bound}) we obtain
\begin{eqnarray}
\nonumber \partial_t s & = & -\gamma_\mathrm{s} s + \mi \sqrt{\frac{d\gamma}{\tau}}\frac{\Omega}{\Gamma_\mathrm{s}} b + e^{-\mi k_\mathrm{s} L}\frac{t_r}{r\sqrt{\tau}} S_\mathrm{in}, \\
\nonumber \partial_t a & = & -\gamma_\mathrm{a} a  + \mi \sqrt{\frac{d\gamma}{\tau}}\frac{\Omega}{\Gamma_\mathrm{a}} b^\dagger + e^{-\mi k_\mathrm{a} L}\frac{t_r}{r\sqrt{\tau}} A_\mathrm{in}, \\
\label{intra} \partial_t b & = &  \mi \sqrt{\frac{d\gamma}{\tau}}\left[ -\frac{\Omega^*}{\Gamma_\mathrm{s}} s + \frac{\Omega}{\Gamma_\mathrm{a}}a^\dagger \right] - \left[ \frac{1}{\Gamma_\mathrm{s}}  + \frac{1}{\Gamma_\mathrm{a}^*} \right] |\Omega|^2 b,
\end{eqnarray}
where we have defined the intra-cavity field-amplitudes $a = \sqrt{\tau}A_0$, $s=\sqrt{\tau}S_0$, as in \cite{Raymer:2013aa}, and similarly $b = \sqrt{L}B_0$, and the resonant and anti-resonant decay rates
\begin{equation}
\label{decay} \gamma_\mathrm{s,a} = e^{-\mi k_\mathrm{s,a} L}\frac{1-\mu_\mathrm{s,a} e^{\mi \phi_\mathrm{s,a}}}{r\tau},
\end{equation}
where $\phi_\mathrm{s,a} = k_\mathrm{s,a}L - \Im\{\kappa_\mathrm{s,a}\}\tau$ is the cavity roundtrip phase accumulated by the fields, including any dispersion induced by the atomic ensemble, and where the cavity roundtrip amplitude transmission, including any atomic absorption, is given by $\mu_\mathrm{s,a} = r e^{-\Re\{\kappa_\mathrm{s,a}\}\tau}$. If there are additional losses due to scattering from surfaces inside the cavity, or partial transmission through cavity mirrors, these can be incorporated into $\mu_\mathrm{s,a}$. The system of coupled equations (\ref{intra}) takes a form that might be written down using cavity input-output theory, except that we are able to treat the dynamics of both the Stokes and anti-Stokes fields, with one being resonant and the other off-resonant with the cavity, as determined by the phases $\phi_\mathrm{s,a}$.

\section{Bad cavity limit}
To proceed to a solution, we once more invoke the bad cavity approximation, in which the fields impinging on the cavity are much more narrowband than the cavity linewidth \cite{Gorshkov:2007aa}. Specifically, then, we require that $|\partial_t s| \ll |\gamma_\mathrm{s} s|$, with commensurate bandwidths for the anti-Stokes and control fields. In this limit, we solve for $s$ and $a$ in terms of $b$ by setting $\partial_t s = \partial_t a \approx 0$, to obtain
\begin{eqnarray}
\nonumber s&=&\mi \sqrt{\frac{d\gamma}{\tau}}\frac{\Omega}{\Gamma_\mathrm{s} \gamma_\mathrm{s}} b+e^{-\mi k_\mathrm{s} L} \frac{t_r}{\gamma_\mathrm{s} r \sqrt{\tau}}S_\mathrm{in}, \\
\label{dtzero} a&=&\mi \sqrt{\frac{d\gamma}{\tau}}\frac{\Omega}{\Gamma_\mathrm{a} \gamma_\mathrm{a}} b^\dagger +e^{-\mi k_\mathrm{a} L} \frac{t_r}{\gamma_\mathrm{a} r \sqrt{\tau}}A_\mathrm{in}, \\
\nonumber
 \partial_t b &=&  \left\{\frac{d\gamma}{\tau}\left[ \frac{1}{\Gamma_\mathrm{s}^2 \gamma_\mathrm{s}} + \frac{1}{|\Gamma_\mathrm{a}|^2 \gamma_\mathrm{a}^*}\right] -  \frac{1}{\Gamma_\mathrm{s}}  -  \frac{1}{\Gamma_\mathrm{a}^*} \right\}|\Omega|^2 b \\
\nonumber & & +\mi \frac{t_r\sqrt{d\gamma}}{r\tau}\left[ \frac{\Omega^*}{\Gamma_\mathrm{s} \gamma_\mathrm{s}} e^{-\mi k_\mathrm{s} L}S_\mathrm{in} +\frac{\Omega}{\Gamma_\mathrm{a} \gamma_\mathrm{a}^*}e^{\mi k_\mathrm{a} L}A_\mathrm{in}^\dagger\right].
\end{eqnarray}
At this point, it is convenient to remove the dependence on the temporal shape of the control field by making a coordinate transformation $t \longrightarrow \epsilon(t) = \int_{-\infty}^t |\Omega(t')|^2\,\rmd t'/W$, with $W$ chosen so that $\epsilon(\infty) = 1$. Then we have $\partial_t = W^{-1}|\Omega(t)|^2 \partial_\epsilon$ \cite{Nunn:2007aa, Reim:2010kx}. Defining normalised signal and anti-Stokes field amplitudes $\sigma = s\sqrt{W}/\Omega$ and $\alpha = a\sqrt{W}/\Omega$, the system (\ref{dtzero}) becomes
\begin{eqnarray}
\nonumber \sigma&=&c_\mathrm{s} b +p_\mathrm{s}\sigma_\mathrm{in}, \\
\nonumber \alpha&=&c_\mathrm{a}b^\dagger +p_\mathrm{a}\alpha_\mathrm{in}, \\
\nonumber
 \partial_\epsilon b &=& f b + g_\mathrm{s} \sigma_\mathrm{in} + g_\mathrm{a} \alpha_\mathrm{in}^\dagger,
\end{eqnarray}
where we pulled the various constants into the coefficients
\begin{eqnarray}
\nonumber c_\mathrm{s,a} &=& \mi \sqrt{\frac{d\gamma W}{\tau}}\frac{1}{\Gamma_\mathrm{s,a} \gamma_\mathrm{s,a}},\\
\nonumber  p_\mathrm{s,a} &=& \frac{t_re^{-\mi k_\mathrm{s,a} L}}{\gamma_\mathrm{s,a}r\sqrt{\tau}},\\
\label{defs} f &=&W\left[\frac{d\gamma}{\tau} \left(\frac{1}{\Gamma_\mathrm{s}^2 \gamma_\mathrm{s}} + \frac{1}{|\Gamma_\mathrm{a}|^2\gamma_\mathrm{a}^*}\right) - \frac{1}{\Gamma_\mathrm{s}} - \frac{1}{\Gamma_\mathrm{a}^*} \right],\\
\nonumber g_\mathrm{s} &=& \mi \frac{t_r e^{-\mi k_\mathrm{s} L} \sqrt{d\gamma W}}{\Gamma_\mathrm{s} \gamma_\mathrm{s}\tau r}\,;\quad g_\mathrm{a} = \mi \frac{t_r e^{\mi k_\mathrm{a} L} \sqrt{d\gamma W}}{\Gamma_\mathrm{a} \gamma_\mathrm{a}^*\tau r}.
\end{eqnarray}
Note that the above transformations are unitary so the operators $\alpha$, $\sigma$, $b$ obey the canonical relations $[\alpha,\alpha^\dagger]=[\sigma,\sigma^\dagger]=[b,b^\dagger]=1$. Solving for $b$ gives
\begin{eqnarray}
\nonumber b(\epsilon) &=& b(0)e^{f\epsilon} + \int_0^1\!\!\rmd \epsilon' \, M_\mathrm{c}(\epsilon,\epsilon')\Big\{ g_\mathrm{s} \sigma_\mathrm{in}(\epsilon') \\
\label{bsolve} & &\left.+ g_\mathrm{a} \alpha_\mathrm{in}^\dagger(\epsilon')\right\},
\end{eqnarray}
where we have defined $M_\mathrm{c}(\epsilon,\epsilon') = \Theta(\epsilon-\epsilon')M(\epsilon,\epsilon')$ as the causal version of the cavity response function $M(\epsilon,\epsilon') = e^{f[\epsilon-\epsilon']}$, with $\Theta$ denoting the Heaviside step-function.

Now we are ready to solve for the outgoing fields emerging from the cavity, using again the beam splitter boundary conditions at the input-output coupler,
\begin{eqnarray}
\nonumber S_\mathrm{out}&=& t_re^{\mi k_\mathrm{s}L} S_L-rS_\mathrm{in}(t),  \\
\label{out} A_\mathrm{out}&=& t_re^{\mi k_\mathrm{a}L} A_L-rA_\mathrm{in}(t).
\end{eqnarray}
In the limit of weak single-pass coupling, we can approximate (\ref{Taylor}) by $e^{\mi k_\mathrm{s} L}S_L \approx e^{\mi \phi_\mathrm{s}}S_0$, retaining only the phase evolution of the Stokes field as it traverses the cavity \footnote{Note that re-using the Taylor expansion in (\ref{Taylor}) actually exacerbates the truncation error of the weak-interaction approximation and gives an incorrect result for (\ref{outfield}).}.  Substituting this into (\ref{out}) and using (\ref{dtzero}), the output field, switching back to the normalised variables $\sigma_\mathrm{in,out} = S_\mathrm{in,out}\sqrt{W}/\Omega$, is found to be
\begin{equation}
\label{outfield}
\sigma_\mathrm{out} = \frac{\mi t_r e^{\mi \phi_\mathrm{s}}\sqrt{d\gamma W}}{\Gamma_\mathrm{s} \gamma_\mathrm{s}\tau}b +\left(\chi e^{\mi \phi_\mathrm{s}} -r\right)\sigma_\mathrm{in},
\end{equation}
where we have defined the cavity transmission amplitude
$$
\chi = \frac{t_r^2}{1-\mu_\mathrm{s} e^{\mi \phi_\mathrm{s}}}.
$$
\section{Storage and retrieval}
We consider first the field emerging from the cavity when we attempt to store an incident signal. This field will contain both a contribution from the un-stored signal due to the finite efficiency of the storage interaction, and also a contribution from four-wave mixing noise. Assuming for simplicity that the memory is initially prepared with no spin wave excitations \footnote{For memories based on alkali vapours --- Rb or Cs --- this requires efficient optical pumping of the atomic ensemble into one of the two hyperfine states within its ground state manifold. In practice this is hard to achieve, but fortunately experiments have shown that a small residual population of un-pumped atoms does not significantly affect the operation of the memory \cite{Michelberger:2015aa}.}, the output field is found from (\ref{outfield}) and (\ref{bsolve}) to be
\begin{eqnarray}
\nonumber \sigma_{\mathrm{out},1}(\epsilon) &=& \int_0^1 \!\!\rmd\epsilon' \, \Big\{ M_1(\epsilon,\epsilon')\sigma_{\mathrm{in},1}(\epsilon') \\
\label{out1}& & \left. + M_\mathrm{FWM}^{[1]}(\epsilon,\epsilon')\alpha_{\mathrm{in},1}^\dagger(\epsilon')\right\},
\end{eqnarray}
where we have defined the integral kernels
\begin{eqnarray}
\nonumber M_1(\epsilon,\epsilon') &=& -r e^{\mi(\phi_\mathrm{s}+k_\mathrm{s}L)} \chi C_\mathrm{s}^2M_\mathrm{c}(\epsilon,\epsilon') \\
\nonumber & &+\left(\chi e^{\mi \phi_\mathrm{s}}- r\right) \delta(\epsilon-\epsilon');\\
\label{kernels1} M_\mathrm{FWM}^{[1]}(\epsilon,\epsilon') &=& -r\chi  e^{\mi(\phi_\mathrm{s}+k_\mathrm{s}L)}C_\mathrm{s}C_\mathrm{a}xM_\mathrm{c}(\epsilon,\epsilon'),
\end{eqnarray}
where the subscript $1$ indicates fields associated with the storage interaction.
Here we have introduced the signal and anti-Stokes memory coupling parameters
\begin{equation}
\label{Cs}
C_\mathrm{s,a} = \frac{\sqrt{\mathcal{C}\gamma W}}{\Gamma_\mathrm{s,a}},
\end{equation}
with $\mathcal{C} = d/(1-\mu_\mathrm{s}e^{\mi \phi_\mathrm{s}})$ the cooperativity of the cavity for the signal field, roughly equal to the optical depth of a medium with length $L\times \mathcal{F}_\mathrm{s}/\pi$, where $\mathcal{F}_\mathrm{s}$ is the cavity finesse for the signal field (when the control field $\Omega$ is not present). Note that the coupling of the anti-Stokes field is multiplied by the \emph{noise suppression factor}
\begin{equation}
\label{xdef}
x = \frac{e^{\mi k_\mathrm{s}L}\gamma_\mathrm{s}}{e^{-\mi k_\mathrm{a}L}\gamma_\mathrm{a}^*} = \frac{1-\mu_\mathrm{s}e^{\mi \phi_\mathrm{s}}}{1-\mu_\mathrm{a}e^{-\mi \phi_\mathrm{a}}}.
\end{equation}
This is the factor that we will aim to minimise by appropriate tuning of the cavity resonances.

We next consider the retrieval interaction. In this case there is no incident signal field, and the initial spin wave is (neglecting decoherence during storage) given by the excitation generated at read-in, so that the retrieved signal field can be written as
\begin{eqnarray}
\nonumber \sigma_{\mathrm{out},2}(\epsilon) &=&\int_0^1 \!\!\rmd\epsilon' \, \Big\{ M_2(\epsilon,\epsilon')\sigma_{\mathrm{in},1}(\epsilon') \\
\nonumber & &  + M_\mathrm{FWM}^{[2]}(\epsilon,\epsilon')\alpha_{\mathrm{in},1}^\dagger(\epsilon')\\
\label{out1} & & \left. + M_\mathrm{FWM}^{[1]}(\epsilon,\epsilon')\alpha_{\mathrm{in},2}^\dagger(\epsilon')\right\},
\end{eqnarray}
where we have defined the integral kernels
\begin{eqnarray}
\nonumber M_2 &=& -\chi r  e^{\mi(\phi_\mathrm{s}+k_\mathrm{s}L)}C_\mathrm{s}^2e^f M;\\
\label{kernels2} M_\mathrm{FWM}^{[2]} &=& -\chi r e^{\mi(\phi_\mathrm{s}+k_\mathrm{s}L)} C_\mathrm{s}C_\mathrm{a}xe^f M.
\end{eqnarray}
Here we have assume for convenience in simplifying the expressions that the control pulse used to drive the retrieval interaction is identical to the storage control pulse.

\section{Efficiency and noise}
The above expressions provide a means to predict the efficiency of the memory by comparing the expectation values $N_\mathrm{x} = \langle \int_0^1 S_\mathrm{x}^\dagger(t)S_\mathrm{x}(t)\,\rmd t \rangle = \langle \int_0^1 \sigma_\mathrm{x}^\dagger(\epsilon)\sigma_\mathrm{x}(\epsilon)\,\rmd \epsilon \rangle$ of the input and output intensity operators ($\mathrm{x}=\mathrm{in,out},1,2$). Four-wave mixing noise enters via the operators describing the incident anti-Stokes fields, which, although they are in the vacuum state (we do not send in any anti-Stokes light), nonetheless contribute to the intensity because the incident fields are described by creation operators, giving rise to anti-normally-ordered terms. In a noiseless memory one would define the total efficiency $\eta_\mathrm{tot, noiseless} = N_{\mathrm{out},2}/N_{\mathrm{in},1}$, with commensurate definitions for the storage efficiency $\eta_\mathrm{store, noiseless} = 1-N_{\mathrm{out},1}/N_{\mathrm{in},1}$ and the retrieval efficiency $\eta_\mathrm{ret, noiseless} = \eta_\mathrm{tot, noiseless}/\eta_\mathrm{store, noiseless}$. For a noisy memory we modify the definition of the total efficiency by subtracting the noise floor $N_{\mathrm{out},2}|_\mathrm{no\;input}$ retrieved from the memory when $N_{\mathrm{in},1} = 0$ (no incident signal photons):
\begin{equation}
\label{etatot}
\eta_\mathrm{tot} = \frac{\widetilde{N}_{\mathrm{out},2}}{N_{\mathrm{in},1}}\,;\qquad \widetilde{N}_{\mathrm{out},2} = N_{\mathrm{out},2}-N_{\mathrm{out},2}|_\mathrm{no\;input}.
\end{equation}
In computing these expectation values, it is helpful to adopt matrix notation for the integral kernels, so that for a two-dimensional function $K = K(\epsilon,\epsilon')$, we can write $\int_0^1 K(\epsilon,\epsilon')K^*(\epsilon'',\epsilon')\,\rmd \epsilon'$ simply as $K K^\dagger = KK^\dagger(\epsilon,\epsilon'')$, and $\int_0^1 K(\epsilon,\epsilon')\psi(\epsilon')\,\rmd\epsilon'$ simply as $K\ket{\!\psi}$, where the ket notation is unrelated to the quantum mechanics of the problem, but is used to denote the vectorised version of the function $\psi(\epsilon)$. In this notation the trace is given by $\mathrm{tr}\left\{K\right\} = \int_0^1 K(\epsilon,\epsilon)\,\rmd \epsilon$. With these preliminaries, we can write the transmitted and retrieved photon numbers, for the case of input coherent state or Fock state signal fields, as
$$
N_{\mathrm{out},j} = \mathrm{tr}\left\{ P_j\right\},
$$
where we have defined the operators
\begin{equation}
\label{Ps} P_j = N_\mathrm{in,1}\ket{\varphi_j}\!\!\bra{\varphi_j} + \sum_{k=1}^jM_{\mathrm{FWM}}^{[k]}M_{\mathrm{FWM}}^{[k]\dagger},\\
\end{equation}
with $\ket{\varphi_j} = M_j\ket{\psi_\mathrm{in}}$, and $\psi_\mathrm{in}(\epsilon)$ the mode-function describing the temporal amplitude of the input signal field, normalised so that $\braket{\psi_\mathrm{in}}{\psi_\mathrm{in}}=1$. To see how this works, consider the temporal amplitude of the signal field emerging from the storage interaction,
$$
\varphi_1(\epsilon) = -r e^{\mi(\phi_\mathrm{s}+k_\mathrm{s}L)} \chi C_\mathrm{s}^2e^{f\epsilon}\psi(\epsilon) + \left(\chi e^{\mi \phi_\mathrm{s}} -r\right) \psi_\mathrm{in}(\epsilon),
$$
where $\psi(\epsilon) = \int_0^\epsilon e^{-f\epsilon'}\psi_\mathrm{in}(\epsilon')\,\rmd\epsilon'$. The number of photons, including noise due to four-wave mixing, is then found to be
$$
N_{\mathrm{out},1} = N_{\mathrm{in},1}\braket{\varphi_1}{\varphi_1} + |r\chi C_\mathrm{s}C_\mathrm{a}x|^2\frac{1-E}{\zeta},
$$
where we have defined $E = \int_0^1 e^{-\zeta \epsilon}\,\rmd \epsilon =  (1-e^{-\zeta})/\zeta$, with the dimensionless coupling parameter $\zeta$ given by
\begin{eqnarray}
\nonumber \zeta &=& -(f+f^*)\\
\nonumber &=&-2r\Re\left\{ C_\mathrm{s}^2 e^{\mi k_\mathrm{s}L} + C_\mathrm{a}^{2} x e^{-\mi (k_\mathrm{a}L - 2\arg \Gamma_\mathrm{a})}
\right\}\\
\label{zetadef} & & + 2W\Re\left\{\frac{1}{\Gamma_\mathrm{s}} + \frac{1}{\Gamma_\mathrm{a}^*}\right\}.
\end{eqnarray}
Considering now the retrieval interaction, the temporal mode emerging from the memory is
$$
\varphi_2(\epsilon) = -r e^{\mi(\phi_\mathrm{s}+k_\mathrm{s}L)}\chi C_\mathrm{s}^2 e^f (e^\zeta E)^{1/2}\kappa e^{f\epsilon},
$$
where we have introduced the normalised overlap between the input field and the cavity response, $\kappa = (e^\zeta E)^{-1/2} \psi(1)$. With this definition, when $\psi_\mathrm{in}(\epsilon) \propto e^{f^*\epsilon}$, we obtain $\kappa = 1$. Including the contributions from four-wave mixing, the number of photons retrieved from the memory is found to be
\begin{equation}
\label{Nout2}
N_{\mathrm{out},2} = |r\chi C_\mathrm{s}|^2 \left[|C_\mathrm{s}E\kappa|^2  N_{\mathrm{in},1} + |C_\mathrm{a}x|^2g(\zeta)\right],
\end{equation}
where we defined $g(\zeta)= (1-e^{-\zeta}E)/\zeta$. The first term describes the coherent operation of the memory; the second term describes the retrieval of noise photons that are present even when no signal photons are sent into the memory. The efficiency of the memory is seen to be
\begin{equation}
\label{eff_1}
\eta_\mathrm{tot} = |r\chi C_\mathrm{s}^2E \kappa|^2.
\end{equation}
Taking the ratio of the first and second terms in (\ref{Nout2}) provides the following formula for the signal-to-noise ratio (SNR) of the memory,
\begin{equation}
\label{SNR}
\mathrm{SNR} = N_{\mathrm{in},1}|\kappa|^2\times \left|\frac{\Gamma_\mathrm{a}}{\Gamma_\mathrm{s}}\right|^2\times \frac{E^2}{g(\zeta)}\times \frac{1}{|x|^2}.
\end{equation}
Parsing this result from left to right, the SNR increases with the number of incident signal photons, and with the degree to which they overlap with the cavity response, described by $|\kappa|^2$. The second factor is essentially the ratio of the anti-Stokes and Stokes detunings, which factor is also present in a cavity-less $\Lambda$-memory. This reflects the fact that low-noise operation can be achieved with detunings much smaller than the splitting between the ground states of the $\Lambda$-system, as can be realised with EIT in cold atoms \cite{Chen:2013aa}. The third factor is purely dynamical, but the final factor, proportional to $|x|^{-2}$, represents the noise suppression afforded by the cavity. As should now be clear, minimising $|x|$ by appropriate tuning of the cavity resonances provides a route to low-noise operation of a $\Lambda$-memory, even at room temperature where large detunings from resonance are necessary.
\section{Mode selectivity}
The cavity memory interaction is a single-mode interaction. That is to say, the memory stores just a single temporal mode, and a single temporal mode is retrieved from the memory \cite{Wasilewski:2009aa}. This is clear from the structure of the solution (\ref{out1}), where the coherent mapping between input and output is described by the Green's function $M_2 \propto M=e^{f[\epsilon-\epsilon']}$, which is a \emph{separable} function of $\epsilon$ and $\epsilon'$. Accordingly the efficiency of the memory is parameterised by the overlap integral $\kappa$ defined above; any input mode orthogonal to $e^{-f\epsilon'}$ will not couple to the memory at all. Furthermore, converting back into the ordinary time coordinate, $\psi_\mathrm{in}(t) =W^{-1/2} \Omega(t) \psi_\mathrm{in}[ t(\epsilon) ]$, we observe that the input mode that is stored can be arbitrarily chosen by appropriate shaping of the control field $\Omega$. The single-mode nature of the cavity memory interaction has been derived previously, albeit without considering four-wave mixing \cite{Cirac:2004aa,Gorshkov:2007aa}, but here we point out that it is an advantageous feature. The combination of single-mode operation and arbitrary shaping has been dubbed \emph{temporal mode selectivity}, and is a useful feature for quantum optical information processing, where large-alphabet signals can be demultiplexed by a mode-selective `drop filter' \cite{Reddy:2013aa}. In fact, the conventional traveling-wave Raman protocol \cite{Nunn:2007aa, Gorshkov:2007qm} is nearly single-mode \cite{Nunn:2008oq}, but the mode selectivity degrades at high efficiency --- a phenomenon encountered also in engineering mode-selective frequency conversion \cite{Eckstein:2011aa}. Recently Reddy \emph{et al.} showed how to achieve high mode selectivity and high efficiency in frequency conversion by double-passing or multi-passing the active medium, so that each interaction was weak enough to remain effectively single-mode \cite{Reddy:2013aa}. The perfect mode-selectivity of the cavity memory analysed here can be understood as the multi-pass limit of this approach, where the interaction describing a single pass through the cavity is weak, and therefore separable, whereas the coherent combination of all cavity round-trips provides for arbitrarily high efficiency while retaining temporal mode selectivity.


\section{Autocorrelation}
A key figure of merit for the operation of a quantum memory is the ability to preserve the sub-Poissonian statistics of stored single-photon Fock states. In our recent experiments with a Raman $\Lambda$-memory without any cavity, we found that although the average number of noise photons generated by four-wave mixing in the absence of an input signal field was low, the effect of four-wave mixing on the photon statistics of the fields retrieved from the memory was dramatic \cite{Michelberger:2015aa}. To see how the cavity influences the performance of the memory as a component for synchronising photonic quantum information, we compute the $g^{(2)}$ autocorrelation function for the fields emerging from the memory, which can be written as
\begin{eqnarray}
\nonumber
g^{(2)}_{\mathrm{out},j} &=& \frac{ \int_0^1\!\! \int_0^1 \!\! \rmd \epsilon \rmd \epsilon' \left \langle  \sigma^\dagger_{\mathrm{out},j}(\epsilon)\sigma_{\mathrm{out},j}^\dagger(\epsilon') \sigma_{\mathrm{out},j}(\epsilon')\sigma_{\mathrm{out},j}(\epsilon) \right \rangle}{\left[ \int_0^1 \!\!\rmd \epsilon'' \sigma_{\mathrm{out},j}^\dagger(\epsilon'') \sigma_{\mathrm{out},j}(\epsilon'')\right]^2}\\
\label{g2}&=& 1 + \frac{ \mathrm{tr}\left\{P_j^2\right\}  - \left(2-g^{(2)}_\mathrm{in,1}\right)N_{\mathrm{in},1}^2\braket{\varphi_j}{\varphi_j}^2}{N_{\mathrm{out},j}^2}.
\end{eqnarray}
For the fields emerging from the storage interaction, we obtain the result
\begin{eqnarray}
\label{g2out} N_{\mathrm{out},1}^2\left[ g^{(2)}_{\mathrm{out},1}-1\right] &=&  \left[g^{(2)}_{\mathrm{in},1}-1\right]N_{\mathrm{in},1}^2 \braket{\varphi_1}{\varphi_1}^2 \\
\nonumber & & + \mathrm{tr}\left\{ M_\mathrm{FWM}^{[1]} M_\mathrm{FWM}^{[1]\dagger} M_\mathrm{FWM}^{[1]} M_\mathrm{FWM}^{[1]\dagger} \right\}\\
\nonumber & &  + 2 N_{\mathrm{in},1} \bra{\varphi_1} M_\mathrm{FWM}^{[1]} M_\mathrm{FWM}^{[1]\dagger} \ket{\varphi_1},
\end{eqnarray}
where the second term evaluates to
\begin{eqnarray}
\nonumber \mathrm{tr}\left\{ M_\mathrm{FWM}^{[1]} M_\mathrm{FWM}^{[1]\dagger} M_\mathrm{FWM}^{[1]} M_\mathrm{FWM}^{[1]\dagger} \right\} &=& |r\chi C_\mathrm{s}C_\mathrm{a}x|^4\zeta^{-4}\times \\
\nonumber & & \left\{  4\zeta (1-E)\right.\\
\label{trM}& & \left. -2\zeta^2 E -\Sigma\right\},
\end{eqnarray}
with $\Sigma = 2e^{-\zeta}\left[\sinh(\zeta)-\zeta\right]$. Of particular interest is the corresponding expression for the $g^{(2)}$ autocorrelation of the fields retrieved from the memory by the readout control pulse, which, after some rather lengthy calculations can be written as,
\begin{eqnarray}
\nonumber N_{\mathrm{out},2}^2\left[ g^{(2)}_{\mathrm{out},2}-1\right] &=& |r\chi C_\mathrm{s}|^4\bigg\{ \left[g^{(2)}_{\mathrm{in},1}-1\right]N_{\mathrm{in},1}^2 |C_\mathrm{s} E \kappa |^4 \\
\label{g2out} & & + |C_\mathrm{a}x|^4h(\zeta)\\
\nonumber & &  + 2 N_{\mathrm{in},1} |C_\mathrm{s}C_\mathrm{a}x\kappa|^2 h'(\zeta)\bigg\},
\end{eqnarray}
where we defined the functions
\begin{eqnarray}
\nonumber h(\zeta)&=& 4\frac{1-E}{\zeta^3} -2\frac{E}{\zeta^2} + E^4  + \frac{\Sigma}{\zeta^4}(2\zeta E-1),\\
\label{hs} h'(\zeta)&=& E^4 + \frac{E\Sigma}{\zeta^3}.\end{eqnarray}
Of particular interest is the case of storing single photon Fock states, for which $g^{(2)}_{\mathrm{in},1} = 0$, and we obtain
\begin{eqnarray}
\nonumber g^{(2)}_{\mathrm{out},2} &=& \left\{2N_{\mathrm{in},1}|\kappa C_\mathrm{s}C_\mathrm{a}x|^2 [E^2g(\zeta) + h'(\zeta)]\right. \\
\nonumber & & \left.+ |C_\mathrm{a}x|^4 \left[ g(\zeta)^2 + h(\zeta)\right]\right\}\\
\nonumber &&\times \left[N_{\mathrm{in},1}|\kappa C_\mathrm{s}E|^2 +|C_\mathrm{a}x|^2 g(\zeta)\right]^{-2}.
\end{eqnarray}
Note that for incident single photons, $N_{\mathrm{in},1}\leq 1$ is to be interpreted as the average number of photons per pulse, including any losses prior to the memory. For a heralded photon source \cite{Spring:2013aa}, $N_{\mathrm{in},1}$ is therefore the \emph{heralding efficiency} of the source. 

If we achieve strong noise suppression, so that $|x|^2\ll N_{\mathrm{in},1}$, then the autocorrelation can be written as
\begin{equation}
\label{g2_low_noise}
g^{(2)}_{\mathrm{out},2} = 2\left[1+\frac{h'(\zeta)}{E^2g(\zeta)}\right]\frac{1}{\mathrm{SNR}}.
\end{equation}

\section{Weak coupling limit}
The expressions for the efficiency, signal-to-noise ratio and autocorrelation simplify in the limit of weak coupling, when $|C_\mathrm{s,a}|\ll 1$, so that $\zeta \ll 1$. In this limit we obtain, to first order in $\zeta$ (valid for $\zeta \lesssim 0.2$), $E\rightarrow 1-\frac{1}{2}\zeta$, $\Sigma \rightarrow \frac{1}{3}\zeta^3-\frac{1}{3}\zeta^4$, $g(\zeta)\rightarrow \frac{3}{2}-\frac{7}{6}\zeta$, $h(\zeta)\rightarrow \frac{11}{6}-\frac{13}{5}\zeta$, $h'(\zeta)\rightarrow \frac{4}{3}
-\frac{5}{2}\zeta$. The number of retrieved photons is then given by
$$
N_{\mathrm{out},2} = |r\chi C_\mathrm{s}|^2\left[N_{\mathrm{in},1}(1-\zeta)|C_\mathrm{s}\kappa|^2 + \left(\frac{3}{2}-\frac{7}{6}\zeta\right)|C_\mathrm{a} x|^2\right],
$$
with the efficiency given by
$$
\eta_\mathrm{tot} = (1-\zeta)|r\chi C_\mathrm{s}^2 \kappa |^2,
$$
and the signal-to-noise ratio
$$
\mathrm{SNR} = \left(\frac{2}{3}-\frac{4}{27}\zeta\right)N_{\mathrm{in},1}|\kappa|^2\left|\frac{\Gamma_\mathrm{a}}{\Gamma_\mathrm{s}}\right|^2\frac{1}{|x|^2}.
$$
The autocorrelation (again for incident single photons in the high noise-suppression regime) is found to be
\begin{equation}
\label{g2_weak}
g^{(2)}_{\mathrm{out},2} = \left(\frac{34}{9}-\frac{14}{81}\zeta\right)\frac{1}{\mathrm{SNR}}.
\end{equation}
In general the control field $\Omega(t)$ can be shaped so as to achieve $|\kappa|=1$ and maximise the SNR. In the weak coupling limit the requirement to shape the signal and control pulses is relaxed, because when $|f|\ll1$ the optimal input mode $e^{f^*\epsilon}$ is approximately flat. If the signal and control pulse shapes derive from a common source, so that $\psi_\mathrm{in}(t)\propto \Omega(t)$ and $\psi_\mathrm{in}(\epsilon)=1$, we then achieve $\kappa = 1-\frac{3}{2}f - \frac{1}{4}f^*\approx 1$.

\section{Strong coupling}
On the other hand it is also instructive to consider the case of very strong coupling, when $\zeta\gg1$. In that case we have $E\rightarrow 1/\zeta$, $g(\zeta)\rightarrow 1/\zeta$, $\Sigma \rightarrow 1$, $h(\zeta)\rightarrow 2(\zeta-1)/\zeta^4$, $h'(\zeta)\rightarrow 2/\zeta^4$, and the memory efficiency tends to
\begin{eqnarray}
\nonumber \eta_\mathrm{tot}&\rightarrow&\left| \frac{r\chi C_\mathrm{s}^2 \kappa }{\zeta}\right|^2  \\
\label{eff_strong} & =&| r\chi \kappa|^2 \frac{|C_\mathrm{s}|^4}{\zeta^2}.
\end{eqnarray}
For very strong coupling, we find $g^{(2)}_{\mathrm{out},2} \longrightarrow 1$, independent of the noise suppression factor $x$. In this case the four-wave mixing gain dominates and the system has effectively become a Raman laser. To remain in the regime where the cavity suppresses the four-wave mixing, we should have $|x|\ll N_{\mathrm{in},1}/\zeta$, in which case from (\ref{g2_low_noise}) we obtain
$$
g^{(2)}_{\mathrm{out},2} = \left|\frac{\Gamma_\mathrm{s}}{\Gamma_\mathrm{a}}\right|^2 \frac{2 |x|^2 \zeta}{N_{\mathrm{in},1}|\kappa|^2}.
$$
Here the autocorrelation rises linearly with the coupling parameter $\zeta$. To achieve low-noise operation of the memory we therefore seek the smallest coupling strength $\zeta$ for which the efficiency saturates.

\section{Comparison with previous results}
Cavity-enhanced $\Lambda$-memories have been analysed previously \cite{Gorshkov:2007aa,Dantan:2005aa,He:2009aa}. To compare with those works, we consider the case that there is no four-wave mixing (equivalent to the limit $\delta \rightarrow \infty$ so that $C_\mathrm{a}=0$), and we assume the signal field to be tuned to the empty-cavity resonance with $k_\mathrm{s}L = 0$~$\mathrm{mod}(2\pi)$. Finally we assume that there are no interface or scattering losses inside the cavity, other than atomic absorption contained in $\kappa_\mathrm{s}$. In this case we can express the cooperativity $\mathcal{C}$ for the signal field in terms of the absorption-free cooperativity $\mathbb{C} = rd/(1-r)$,
\begin{equation}
\label{Cinterms}
\mathcal{C} = \mathbb{C}\frac{1}{r}\frac{\Gamma_\mathrm{s}}{\Gamma_\mathbb{C}},
\end{equation}
where $\Gamma_\mathbb{C} = \Gamma_\mathrm{s} + \mathbb{C}\gamma$. We also find $\zeta  = 2W\gamma (\mathbb{C}+1)/|\Gamma_\mathbb{C}|^2$ and $\chi = (1+r)\Gamma_\mathrm{s}/\Gamma_\mathbb{C}$, which upon substitution into (\ref{eff_strong}) yields the rather elegant result, previously derived by Gorshkov \cite{Gorshkov:2007aa}, that the optimal storage efficiency in the limit of strong coupling is
\begin{equation}
\label{Gorsh_eff}
\eta_\mathrm{tot} = |\kappa|^2\left(\frac{1+r}{2}\right)^2\left[\frac{\mathbb{C}}{\mathbb{C}+1}\right]^2,
\end{equation}
which is independent of the detuning $\Delta_\mathrm{s}$ and also independent of the control field energy $\mathcal{E}\propto W$, provided the control field is strong enough to reach saturation, $\zeta \gg 1$. Unit efficiency is achieved with an optically thick ensemble in a high quality cavity with a mode-matched storage interaction such that $|\kappa|=1$.

In a real atomic system, four-wave mixing is present, and in that case the maximum noise suppression is achieved when the cavity resonance condition $\phi_\mathrm{s} = 0\mod{2\pi}$ obtains, which is different to the empty-cavity resonance condition $k_\mathrm{s}L = 0\mod{2\pi}$ because of atomic dispersion. In fact, the resonance condition $\phi_\mathrm{s} = 0$ is natural in a real experiment, where the atoms are introduced into the cavity before the signal field is tuned to resonance with the cavity. Any other scattering losses inside the cavity should also be included in the roundtrip amplitude transmission $\mu_\mathrm{s}$. Four-wave mixing gain can boost the efficiency (while introducing noise), whereas intra-cavity losses will reduce the efficiency. In general the expression (\ref{eff_1}) and its strong-coupling limit (\ref{eff_strong}) can be used to predict the memory performance.



\section{Double resonance}
We have so far treated the control pulse $\Omega$ as an external parameter, but the control is an optical field that co-propagates with the signal field around the cavity. As depicted in Fig.~\ref{fig:cavity_memory} (b) The control field frequency lies between that of the signal and the anti-Stokes field, so it is not immediately obvious how it can be introduced into the cavity: if the free-spectral range of the cavity is adjusted so that resonance is achieved for the Stokes and anti-resonance is achieved for the anti-Stokes field, separated in frequency by $2\delta$, then the cavity will not be resonant with the control, shifted from the Stokes frequency by $\delta$. In this connection we briefly mention two solutions. First, the control can be polarised orthogonally to the signal field, and a birefringent cavity can be used to shift the cavity resonance for the control polarisation so as to in-couple the control. Second, the atomic dispersion described by the real parts of $\kappa_\mathrm{s,a}$ distorts the regular frequency spacing of the empty-cavity resonances \cite{Munns:2015,Sabooni:2013aa}, and this distortion can be used to bring signal and control (even co-polarised) into simultaneous resonance, while the anti-Stokes is off-resonant. These or other techniques may be used, which lie beyond the scope of this work. In any case, a resonantly coupled control will benefit from intra-cavity field enhancement,
$$
\Omega = \Omega_0\frac{t_\mathrm{\Omega}}{1-\mu_\Omega}\approx \Omega_0 \times \sqrt{\frac{2\mathcal{F}_\mathrm{\Omega}}{\pi}},
$$
where $t_\Omega$, $\mu_\Omega$ are the mirror amplitude transmission and the cavity-roundtrip amplitude transmission for the control mode, with $\mathcal{F}_\mathrm{\Omega}$ the corresponding finesse, and where $\Omega_0$ is the free-space Rabi frequency. Compared to a free-space memory, without any cavity around the atoms, where $\Omega$, $\mathcal{C}$ must be replaced instead by their free-space counterparts $\Omega_0$, $d$, the cavity memory benefits from a double enhancement in coupling strength, by a factor $\sim \mathcal{F}_\mathrm{s}^2\mathcal{F}_\Omega^2/\pi^2$. This doubly-resonant enhancement enables efficient light storage with fewer atoms and much less control pulse energy $\mathcal{E}$ than required to run a free-space memory --- and of course four-wave mixing is suppressed in this configuration, which it is not in a free-space memory.

\section{Bandwidth}
The analysis presented here assumes adiabatic following of the atoms driven by the intra-cavity fields, and also adiabatic evolution of the intra-cavity fields driven by the external fields impinging on the cavity. This \emph{bad cavity} approximation holds when the spectral bandwidth of the external fields is much narrower than the decay rates $\gamma_\mathrm{s,a}$ of the intra-cavity fields. Or equivalently the cavity resonance at the signal frequency, with linewidth $\gamma_\mathrm{s}$ should be broad compared to the signal pulse bandwidth. The acceptance bandwidth $\delta_\mathrm{s}$ of such a memory is therefore limited by the cavity finesse, $\delta_\mathrm{s} < a\Delta_\mathrm{FSR}/\mathcal{F}_\mathrm{s}$, where $a\sim 0.3$ is a `safety margin' and where $\Delta_\mathrm{FSR}$ is the free spectral range of the cavity. Assuming low loss so that $\mu_\mathrm{a}\approx \mu_\mathrm{s}\approx 1$, and the resonance/anti-resonance conditions $\phi_\mathrm{s} = 0\mod{2\pi}$, $\phi_\mathrm{a}=\pi\mod{2\pi}$, we have $x\approx 2\pi/\mathcal{F}_\mathrm{s}$, and so $\delta_\mathrm{s}\propto x$, which expresses a trade-off between noise suppression and speed of operation. Nonetheless we will show in the next section that it is feasible to operate a moderately broadband memory with good noise suppression in alkali vapour.

\section{Cavity-enhanced Raman memory}
\begin{figure}[h]
\begin{center}
\includegraphics[width=10.3cm]{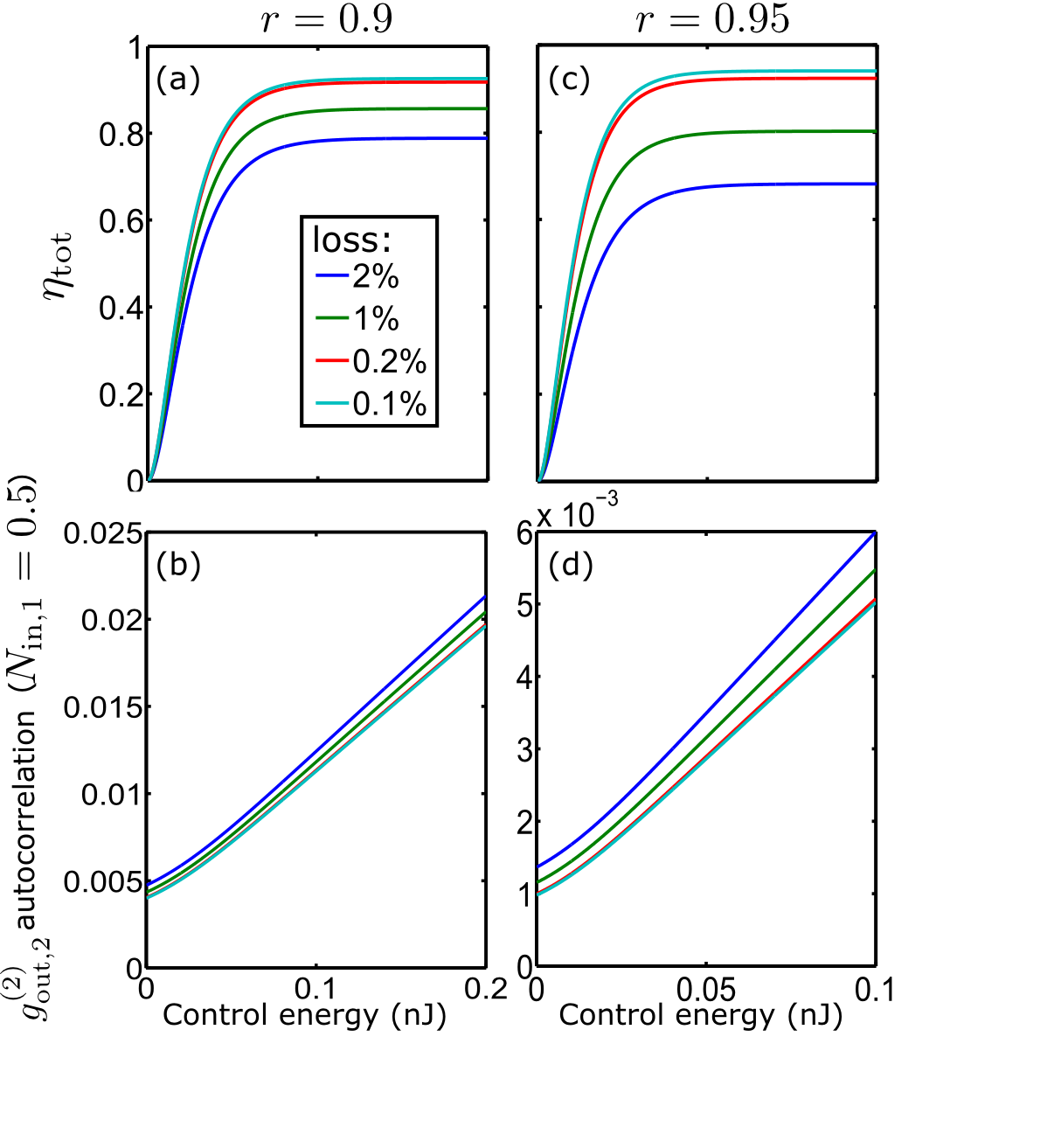}
\vspace{-1.5cm}
\caption{The predicted efficiency $\eta_\mathrm{tot}$ and autocorrelation $g^{(2)}_{\mathrm{out},2}$ for the storage of single photons, heralded with efficiency $N_{\mathrm{in},1} = 0.5$, in a cavity-enhanced Raman memory, as described in the text, with input-output-coupler-reflectivities $r=0.9$ ((a),(b)) and $r=0.95$ ((c),(d)), and a range of intra-cavity roundtrip intensity losses (see legend).}
\label{fig:effs_and_g2s_2}
\end{center}
\end{figure}

Here we consider a specific implementation of a $\Lambda$-memory in caesium vapour, in the far-off-resonant Raman limit. While our results are valid for adiabatic storage with arbitrary detunings, and also on resonance --- which corresponds to EIT-type $\Lambda$-memories --- the Raman configuration has the advantage that broadband pulses can be stored at room temperature outside of the Doppler and collisional absorption profile, and the pulse durations $\sim$ns timescale are sufficiently short that the collisional fluorescence \cite{Manz:2007aa} and dephasing are negligible during the optical interactions \cite{Reim:2011ys}. Preliminary experiments have confirmed the operating principles of the cavity-enhanced memory proposed here \cite{Saunders:2015aa}, and this calculation indicates that a high-performance memory is within reach. We begin by noting the caesium hyperfine splitting of $\delta=9.2$~GHz $=57.8\times 10^9$s$^{-1}$. For simplicity consider the proposal, mentioned above, to use a tunable birefringence to in-couple an orthogonally-polarised control field. Neglecting for the moment atomic dispersion, this fixes the free-spectral range of the cavity to $\Delta_\mathrm{FSR} = 4\delta/(2m+1)$, where the integer $m=0,1,2...$ specifies the \emph{order} of the cavity --- that is, the number of cavity resonances that lie between the signal and anti-Stokes frequencies. The largest operating bandwidth is achieved for the largest free-spectral range, so we consider a zero-order cavity. The cavity roundtrip length is then fixed to be $L = 2\pi c/\Delta_\mathrm{FSR} = 8$~mm, which is much smaller than a comparable free-space memory and could be incorporated `on chip'. The natural linewidth of the Cs D$_2$ line at $\lambda=852$~nm is $2\gamma = 5.2$~MHz, but with a buffer gas to extend the storage lifetime the pressure-broadened linewidth could be larger, so we will assume $2\gamma \approx 50$~MHz. We then choose a signal detuning of $\Delta_\mathrm{s}=5$~GHz, which puts us in the far-off-resonant limit, and which lies sufficiently outside the Doppler profile, $\sim 500$~MHz at room temperature, that inhomogeneous broadening of the vapour can be neglected. Assuming a temperature of $\sim70^\circ$~C, the optical depth --- fixed by the cavity length $L$ and the temperature-dependence of the caesium vapour pressure --- is found to be $d\approx380$. For a collimated cavity mode the Rayleigh range should be matched to the roundtrip length, which fixes the transverse mode area to $\mathcal{A} = \lambda L$. Figure~\ref{fig:effs_and_g2s_2} shows the efficiency and $g^{(2)}$ autocorrelation function for the storage of single photons heralded with efficiency $N_{\mathrm{in},1}=0.5$ in such a cavity-enhanced Raman memory, as a function of the energy of the control pulses, for two different input-output coupler reflectivities, and assuming several different levels of intra-cavity loss. We consider ideal mode-matching such that the mode overlap is $|\kappa| = 1$. The highest asymptotic efficiencies are achieved for a critically coupled memory that maximises $\chi$, since with strong noise suppression we have $\eta_\mathrm{tot} \leq \chi^2/4$. For an intracavity roundtrip loss parameterised by $\mu_\mathrm{s}/r = \cos^2 \theta \approx 1-\theta^2$ (for $\theta\ll 1$), the optimal reflectivity is $r_\mathrm{opt} = 1-\sqrt{2}\theta$, corresponding to an optimised transmission amplitude $\chi_\mathrm{opt}/2 =  (\sqrt{2}-\sin \theta)/(\sin \theta + \sqrt{2}\cos^2\theta)\approx r_\mathrm{opt}$. For a lossless cavity we have $r_\mathrm{opt}=\chi_\mathrm{opt}/2 = 1$, but in that case the acceptance bandwidth of the memory is zero. In general, then, after minimising intra-cavity losses, the input-output coupler reflectivity should be maximised, subject to achieving the required acceptance bandwidth and noise suppression.

The cavity linewidth is $1.3$~GHz and $0.6$~GHz for the two cases $r = 0.9$, $r = 0.95$, respectively. The memory acceptance bandwidths are then $\delta_\mathrm{s}\sim 400$~MHz, and $200$~MHz, respectively. These are compatible with cavity-enhanced downconversion sources \cite{Brecht:2015ab} and would enable the storage of few-ns-duration pulses, suitable for the fastest useful clock-rates, $\sim$0.1-1~GHz, for electronic synchronisation applications. Finally, the control pulse energies required to saturate the efficiency, on the order of $\mathcal{E}=10$~pJ are sufficiently low that the memory could be driven by the modulated output of a semiconductor diode laser.

\section{Conclusion}
We have described a method to suppress four-wave mixing noise in a $\Lambda$-memory by means of a cavity tuned simultaneously into resonance with the signal field to be stored, and into anti-resonance with the field mode that is excited through spontaneous Raman scattering driven by the control field interacting with the populated initial state. We have extended previous analyses of cavity-enhanced $\Lambda$-memories by explicitly considering the evolution of both the Stokes and anti-Stokes fields, and we have derived analytic formulas for the four-wave mixing noise and --- a key figure of merit for quantum information applications --- the $g^{(2)}$ autocorrelation for the retrieved fields. Our analysis shows that highly efficient and low-noise operation can be simultaneously achieved, and we have presented an example calculation for a memory in Cs vapour which offers the realistic prospect of constructing technically-simple quantum memories for near-infra-red photons, with near-unit efficiency and negligible noise, operated at GHz rates at room temperature in a chip-scale package powered by diode lasers. We are optimistic that technologies of this kind will enable the field of quantum optics to finally escape the `two-photon doldrums' and explore the physics of large quantum systems.

\acknowledgements
JN acknowledges W.~S.~Kolthammer and X.-M.~Jin for early discussions. This work was supported by the UK Engineering and Physical Sciences Research Council (EPSRC) through the standard grant EP/J000051/1, the programme grant EP/K034480/1 and the EPSRC Hub for Networked Quantum Information Technologies (NQIT). Additional support came from the EU IP on Quantum Interfaces, Sensors, and Communication based on Entanglement Integrating Project (Q-ESSENCE; 248095), the Air Force Office of Scientific Research: European Office of Aerospace Research \& Development (AFOSR EOARD; FA8655-09-1-3020), the EU IP SIQS (600645), and EU Marie-Curie Fellowships PIIF-GA-2013-629229 to DJS and PIEF-GA- 2013-627372 to EP. CQ was supported by the China Scholarship Council. JN acknowledges a Royal Society University Research Fellowship. IAW acknowledges an ERC Advanced Grant (MOQUACINO). MGR and DVR acknowledge support from the U.S. National Science Foundation (QIS Program).

\end{document}